# Effect of spatial confinement on spin-wave spectrum: Low temperature deviation from Bloch's $T^{3/2}$ law in Co nanoparticles.

P Anil Kumar* and K Mandal

Magnetism Laboratory, S N Bose National Centre for Basic Sciences,

Block JD, Sector-III, Salt Lake, Kolkata-700 098, India.

**Abstract:**

We present the study on Bloch $T^{3/2}$ law and its applicability in ferromagnetic cobalt nanoparticles with sizes 25 and 38 nm. Bloch has derived the $T^{3/2}$ law by assuming long wave-length spin-waves to be excited at low temperatures. But, in nanoparticles the wavelength of the spin wave is confined by the size of the magnetic particle leading to the gap in the spin-wave energy spectrum. The experimental observation leads to the conclusion that Bloch's law is valid at temperatures higher than the spin-wave energy gap. However, it is not applicable at low enough temperatures, where the energy gap becomes prominent. We have demonstrated that a theory recently developed by us [Mandal et al., Europhys. Lett. **75**, 618 (2006)] explains the variation of magnetization with temperature accurately. In addition, the hysteresis properties of these cobalt nanoparticles are also presented here.

* anil@bose.res.in

## I. Introduction:

The possibility of achieving ultra high-density data storage system using magnetic particles as memory elements and their uses in medicine has recently enhanced the research activities in the area of magnetic nanomaterials[1-3]. These applications need the smallest possible magnetic particles for effective usage. However, as the size of the magnetic grain is reduced below a critical size, the domain wall formation is not supported energetically and hence the single magnetic domain particles are developed. On further reduction of grain size, the anisotropy energy (being proportional to particle volume) holding the particle magnetization in a particular direction becomes comparable to that of the thermal energy at and above superparamagnetic blocking temperature $T_B^{SP}$. This will lead to the phenomenon of superparamagnetism[4]. However, this paper deals with single domain cobalt nanoparticles. The single domain particles vary in their properties from the multi domain counterparts. For example their coercivity is largely increased[5,6], saturation magnetic moment is reduced significantly[5] and the critical temperature of paramagnetic transitions($T_N/T_C$) are varied[7,8]. The absence of domain walls in single domain particles explains the increase in the coercivity. The magnetization/demagnetization takes place by rotation and this requires higher fields compared to the domain wall movement. The reduction in the saturation magnetic moment can be attributed to the existence of large surface spins, which do not contribute same moment as that in core of the particle. And finally the reduction in $T_N/T_C$ is also attributed to the large surface spins which can be deflected at low temperature. There have been studies on the single domain cobalt nanoparticles including the magnetization switching, exchange bias effects, superparamagnetic effect[9-11].

In the present work we study the magnetic behavior of single domain cobalt particles having diameters 25 and 38 nm. The magnetic properties of these particles have been characterized by temperature dependent coercivity, anisotropy, remanance. The applicability of the Bloch's law in these nanoparticles throughout the temperature range has been studied and an explanation for the deviation from the Bloch's law in the low temperature region has been proposed. It is important to note that several researchers have shown the law to be deviated in nanoparticles systems[12-15].

## II. Experimental:

The Co-SiO$_2$ nanocomposite powders were prepared by sol-gel method[16] as described in an earlier work[17]. A clear solution of cobalt nitrate in water was mixed with a solution of TEOS (tetraethoxy orthosilicate) in a mixture of water and ethanol and stirred to mix the solutions well, maintaining a pH close to 2, as it is the isoelectric point. The mixed solution was kept at ambient conditions for 2 days, after which the sol transformed to gel. The so formed gel was reduced in an electric furnace in a continuous flow of H$_2$ gas at various temperatures to obtain samples with different particle sizes of cobalt. We have prepared samples with two different weight (volume) ratios 30:70 (10:90) and 60:40 (27:73) of Co in SiO$_2$. Samples A and B belong to first composition while Sample C belong to the other composition. Sample A is reduced at 550° C for 1 hour, Sample B at 450° C for 1 hour and Sample C at 550° C for 1 hour.

A PANalytical Xpert Pro MPD X-ray diffractometer has been used to study the phase and composition of samples by Cu K$_\alpha$ radiation. A JEOL 2100 model high resolution transmission electron microscope (HRTEM) was used for determining the particle morphology and size. Magnetic measurements were carried out using a Quantum Design MPMS SQUID magnetometer. The saturation magnetization ($M_S$) variation with temperature was measured under the application of a 5T constant magnetic field, from 2K to 310K. The hysteresis loops (figure.3) measured at various temperatures down to 10K indicates the sufficiency of the 5T magnetic field to saturate the sample magnetization.

## III. Results and Discussions:

The X-ray diffraction (XRD) patterns shown in Figure.1 reveal the formation of *hexagonal* Co in all the samples. The particle size has been calculated by substituting the FWHM and θ values of diffraction peaks in Scherrer's formula[18]. The average particle sizes calculated from XRD are 25 nm (samples A and B) and 38 nm (sample C). Transmission electron microscope (TEM) images for sample B are shown in Figure.2 (a). Electron micrographs show the hexagonal faceted cobalt particles of size ~ 29.7 nm. The particle size distribution obtained using the TEM micrograph is shown in Figure.2 (b). The energy dispersive X-ray spectra (EDS) analysis confirmed the presence of cobalt, silicon and oxygen and no other elements in the samples.

Figure 3 shows the hysteresis loops measured at temperatures 10K and 300K for (a) Sample A, (b) Sample B and (c) Sample C. The coercivity of nanoparticles in all the three samples is much higher(~ 870 Oe) compared to coercivity of bulk cobalt (~ 13Oe)[19] and it increases with temperature gradually till 50K and decreases at 10K as detailed in figure 4. The remanence '$M_r$' decreases with increasing temperature shown in figure 4 signifying the increase in thermal oscillations of magnetic moment of the particles. In the high field region the data is analyzed using law of approach to saturation (LAS) as given by equation (1)[20,21].

$$M(H) = M(\infty)(1 - \frac{a}{\sqrt{H}} - \frac{b}{H^2}) \quad (1)$$

The value of uniaxial anisotropy constant $K$ is calculated using the value of '$b$'. It is observed that (from figure.4) the value of $K$ is slightly higher compared to that of the value for bulk cobalt and it gradually decreases with increasing temperature.

Saturation magnetization variation with temperature is presented in figure 5 for samples A, B and C. The nature of the plot is clearly different from that for the bulk Co, which follows Bloch's $T^{3/2}$ law. Usually the demagnetization of ferromagnetic or ferrimagnetic materials at temperatures much below the transition temperature is due to the excitation of long wave-length spin waves whose energy is characterized by spin-wave stiffness coefficient D. The excitation energy $E_k$ of spin wave in the limit of small wave vectors ($ka \ll 1$, where '$a$' is inter-atomic distance) is given by [22]

$$E_k = Dk^2 \quad \ldots\ldots (2)$$

The decrease in magnetization $M_S$ from its saturation value $M_S(0)$ at temperatures well below its critical temperature is determined by the equation [22]

$$M_S(T) = M_S(0) - \int_0^\infty g(k)n(k)dk \ldots (3)$$

Where g(k) is the density of spin-wave states and n(k), the Bose occupation number. Using the Bose-Einstein distribution law and substituting the values of n(k) and g(k), we get,

$$\frac{M_S(0) - M_S(T)}{M_S(0)} = \frac{g\mu_B}{2\pi^2} \int_0^\infty \frac{k^2 dk}{e^{\beta E_K} - 1} \ldots (4)$$

In case of a continuous distribution of spin wave states as in the bulk, the above equation lead to the temperature dependence of magnetization given by the Bloch law as

$$M_S(T) = M_S(0)[1 - BT^{3/2}] \quad \ldots\ldots (5)$$

where $B = 2.6149 V_O \left(\frac{k_B}{4\pi D}\right)^{3/2}$, $V_o$ being the atomic volume. In a recent paper on nickel ferrite nanoparticles[23], we have shown that the Bloch $T^{3/2}$ law given by equation (5) becomes invalid at low temperatures. The Bloch law was derived in the long wavelength limit (of spin-waves) to explain the thermal variation of saturation magnetization in ferromagnetic /ferrimagnetic materials. However, when the particle size of a ferromagnetic or ferrimagnetic sample is reduced to below a few nanometers size, the existence of long wavelength spin-waves becomes questionable. We propose that the *spin-waves cannot have a wavelength greater than double the*

*diameter of magnetic particle.* And this leads to the quantization of spin-wave energy spectrum and thus to a different behavior of magnetization with temperature than predicted by Bloch's law. The finite size of the particles led to a discrete set of energy values corresponding to a discrete spectrum of spin-wave modes. For example, in case of a cubic particle with each side d, the spin wave energies can be roughly estimated as[13]

$$E_n = Dk_n^2 = D(n\pi/d)^2, n = 1, 2, 3, \ldots \quad (6)$$

Therefore the second term in Eq.(2) can not be integrated for all values of k from 0 to ∞ and it may be modified as

$$M_S(T) = M_S(0) - C\sum_{n=1}^{p}\frac{1}{(e^{E_n/kT}-1)} - \int_{k'}^{\infty} g(k)n(k)dk \quad (7)$$

Where $C = M_S(0)/NS$ and the summation runs till 'p' such that $E_p \sim k_BT$. The third term in the above equation can be neglected at low temperatures, as these high-energy modes do not contribute much to the magnetization reduction. Now the term in the summation is modified as

$$\frac{1}{e^{E_n/k_BT}-1} \rightarrow \frac{e^{-E_n/k_BT}}{1-e^{-E_n/k_BT}}$$

Further by using the fact that for $E_n \geq k_BT$, $e^{-E_n/k_BT} < 1$, we can proceed as

$$\frac{e^{-E_n/k_BT}}{1-e^{-E_n/k_BT}} = e^{-E_n/k_BT}[1-e^{-E_n/k_BT}]^{-1} = e^{-E_n/k_BT} + e^{-2E_n/k_BT} + e^{-3E_n/k_BT} + \ldots$$

and the higher power terms are neglected to get

$$\frac{e^{-E_n/k_BT}}{1-e^{-E_n/k_BT}} = e^{-E_n/k_BT}$$

therefore the equation (7) reduces to equation (8) namely

$$M_s(T) = M_s(0) - C\sum_{n=1}^{p} e^{-E_n/k_BT} \quad \ldots (8)$$

The gaps in the spin wave spectrum become ineffective with the increase in thermal energy ($E_n \ll k_BT$). Therefore, at higher temperatures the summation term can be included in the integration term and after integration, Eq.(7) gives an expression similar to the Bloch's $T^{3/2}$ law (Eq.(5)) with the parameters $M_S(0)$ and B having different significance.

Our present experiments further confirm that the Bloch law is indeed invalid at low temperatures and the law [equation (8)], we have derived previously is able to explain the magnetization variation with temperatures. The figure 5 shows the Ms-T curves for the samples, the low temperature data is fitted using equation (8) and the matching is almost exact. However, only a single exponential term in equation (8) is considered for the data of samples A and B whereas up to two exponentials had to be used for fitting the data of sample C. This is because the sample C contains larger particles and they are close compared to those in other samples.

This might lead to the population of higher energy magnon levels. Also the temperature at which the upturn occurs is seen to be lower for sample C. The values of $E_1$ calculated from the fitting of equation (8) to the low temperature data are given in Table I. One can calculate the value of $E_1$ from equation (6) using $k = (2\pi/\lambda) = (\pi/d) \sim \pi/20$ nm$^{-1}$, ('d' being diameter of the particle) and a D value of 600 meV $A^{o2}$. The value of $E_1$ thus calculated turns out to be of the order of meV. The values obtained from the fitting are also of the same order. However, the sample contains particles of distributed diameters and the equation (6) is for a cubic particle. Therefore the value of $E_1$ cannot be matched in exact sense to the gap in spin-wave spectrum.

It should be noted that in a previous study, Eggeman et al. have recently presented similar up turn in the curve of Ms (saturation magnetization) vs T (temperature) in Co nanoparticles[15]. They have, however, attributed this upturn to the surface spin contribution to magnetization. In our case the application of 50 kOe external field is supposed to suppress this or in a way align the surface spins completely. Finally, the issue is still debatable by the researchers and it may have serious implications on the application of magnetic nanoparticles.

## IV. Conclusion:

The main issue of validity of Bloch's $T^{3/2}$ in metallic nanoparticles has been addressed by taking cobalt as the model system. It is observed experimentally that the $T^{3/2}$ law becomes invalid at low temperatures while it is valid at higher enough temperature to overcome the spin-wave energy gap. A law is derived closely following the Bloch's derivation but considering the short-wave length spin-waves. The law thus derived is shown to explain the experimental finding well.

**Acknowledgements:**
One of the authors PAK is thankful to CSIR, Govt of India for a research fellowship. The help of Prof R Mitra and Mr R Basu of IIT Kharagpur is acknowledged in taking HRTEM images. Thanks are due to UNANST, IACS, Kolkata for their support in SQUID measurements.

**Table.I** Description of heat treatment and the values of crystallite size from XRD, B from fitting (of equation (5)) and $E_1$ from fitting of equation (8)

| Sample | Heat treatment in $H_2$ gas | Crystallite size of Co from XRD (St. Dev.) | Value of B from the fit of eqn (5) x $10^{-6}$ ($K^{-3/2}$) | Value of $E_1$ from the fit of eqn (8) (meV) |
|---|---|---|---|---|
| A Series-I | 450°C (1hr) | 24.80nm (2.8) | 9.06 | 2.10 |
| B Series-I | 550°C (1hr) | 24.90nm (2.8) | 9.24 | 1.74 |
| C Series-II | 550°C (1hr) | 37.90nm (3.0) | 8.20 | 0.10 2.79[a] |

[a] value of $E_2$

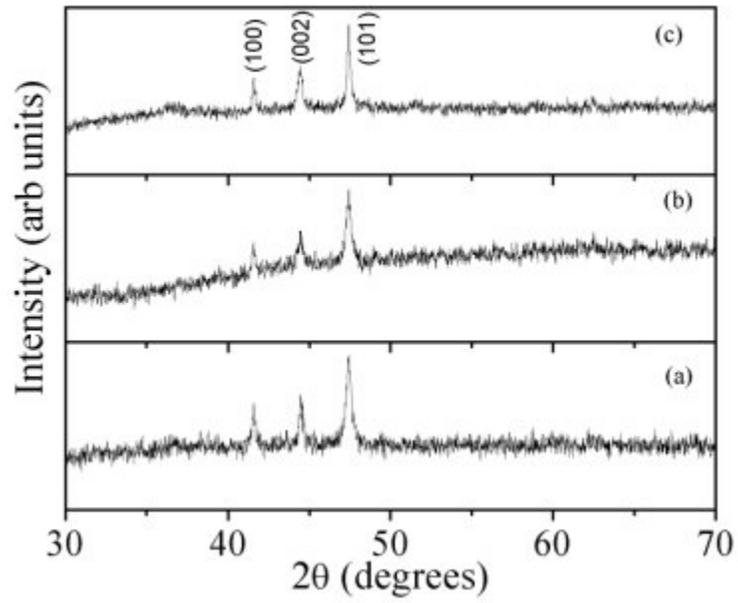

Figure 1: X-ray diffraction pattern for **a)** sample A, **b)** sample B and **c)** Sample C showing the formation of *hexagonal* Co.

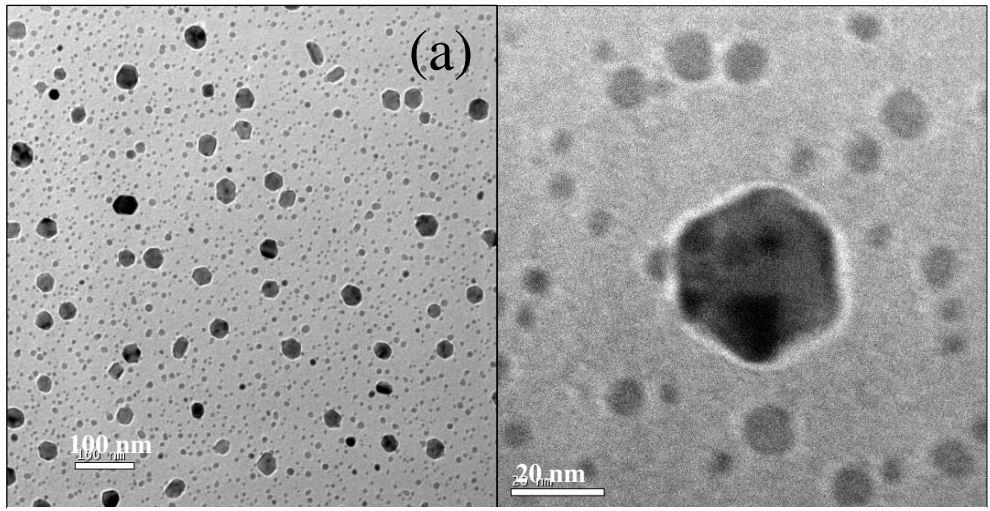

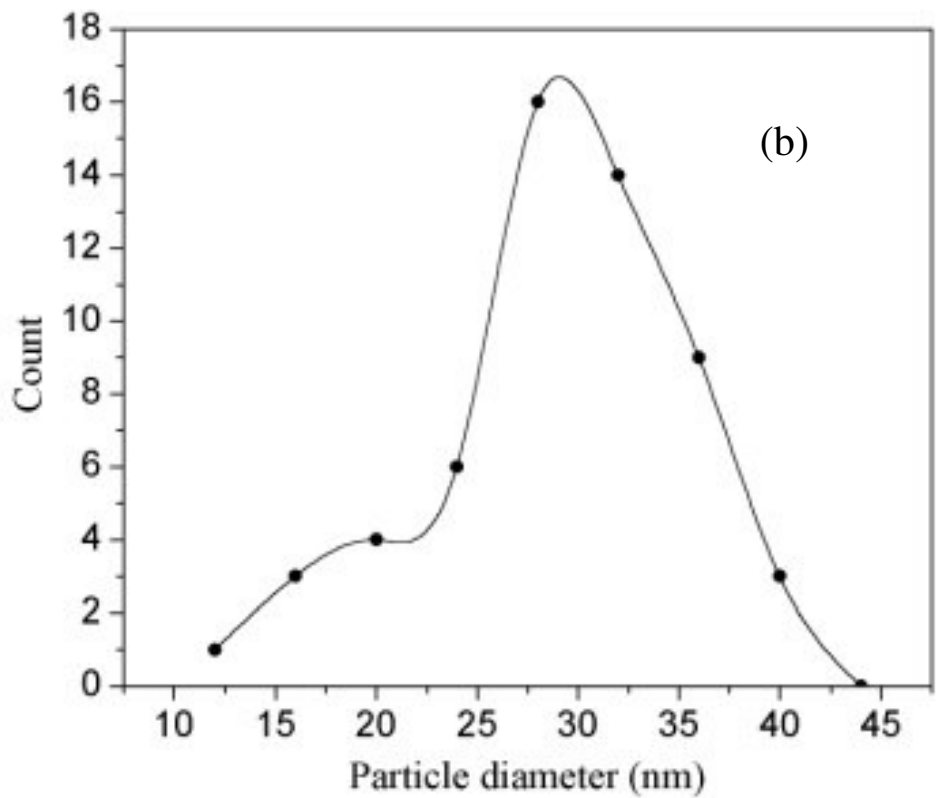

Figure.2: **a)** HRTEM micrographs of sample B **b)** graph showing particle size distribution, line is guide to the eye.

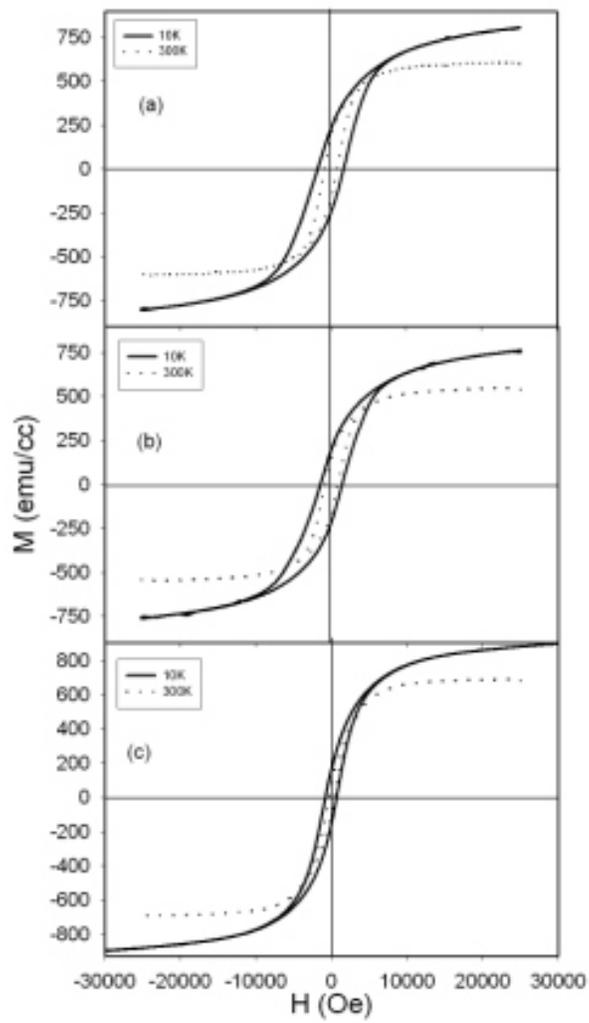

Figure.3: Hysteresis plots of **a)** sample A, **b)** sample B and **c)** sample C at 10K and 300K.

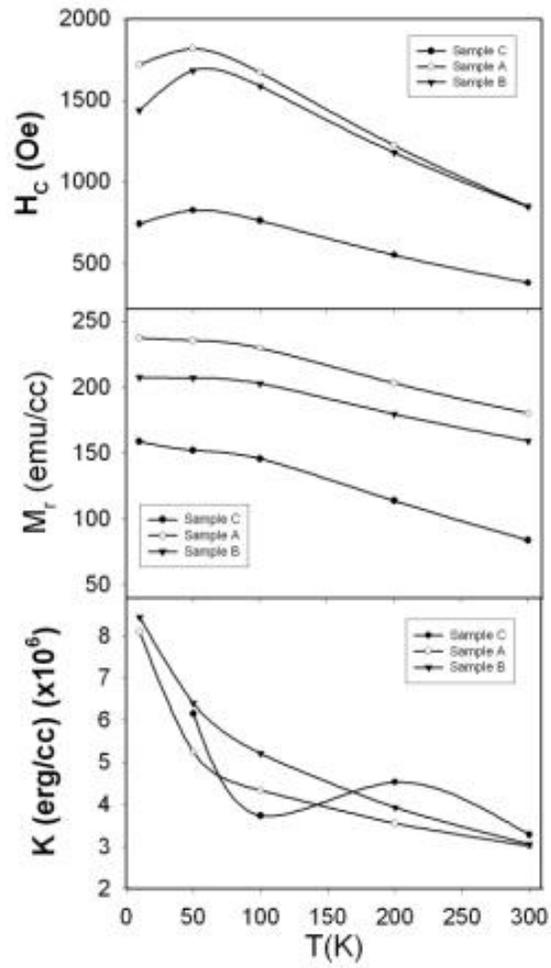

Figure.4: Plot showing the variation of coercivity ($H_C$), anisotropy constant (K) and remanance magnetization ($M_r$) with temperature for the studied samples. Lines are guide to the eye.

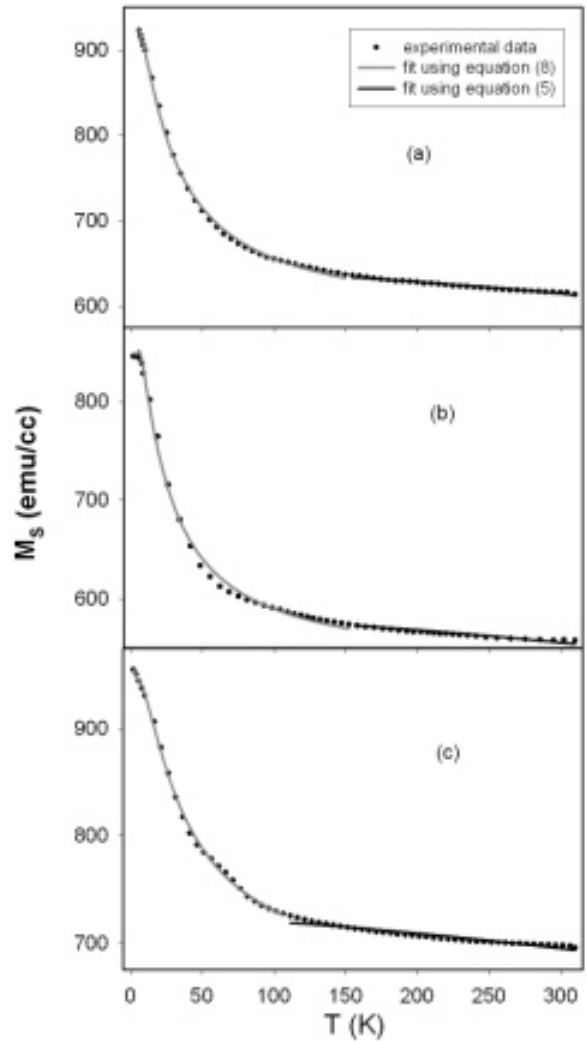

**Figure.5:** Plots showing the variation of Saturation magnetization Ms with temperature for **a)** sample A, **b)** sample B and **c)** sample C. The gray line is a fit to data using equation (8) while the black line is a fit using $T^{3/2}$ law.